\documentclass{ws-ijmpe}

\begin{document}

\markboth{S. Bianchin {\it et al.}}{The HypHI project at GSI and FAIR
}

\catchline{}{}{}{}{}

\title{THE HYPHI PROJECT:\\
HYPERNUCLEAR SPECTROSCOPY WITH STABLE HEAVY ION BEAMS AND RARE ISOTOPE BEAMS AT GSI AND FAIR\\
}

\author{\footnotesize S.~BIANCHIN$^{1}$\footnote{E-mail: s.bianchin@gsi.de}~,
P.~ACHENBACH$^{2}$,
S.~AJIMURA$^{3}$,
O.~BORODINA$^{1,2}$,
T.~FUKUDA$^{4}$,
J.~HOFFMANN$^{1}$,
M.~KAVATSYUK$^{5}$,
K.~KOCH$^{1}$,
T.~KOIKE$^{6}$,
N.~KURZ$^{1}$,
F.~MAAS$^{1,2}$,
S.~MINAMI$^{1,4}$,
Y.~MIZOI$^{4}$,
T.~NAGAE$^{7}$,
D.~NAKAJIMA$^{1,8}$,
A.~OKAMURA$^{6}$,
W.~OTT$^{1}$,
B.~\"OZEL$^{1}$,
J.~POCHODZALLA$^{2}$,
C.~RAPPOLD$^{1,9}$,
T.R.~SAITO$^{1}$,
A.~SAKAGUCHI$^{10}$,
M.~SAKO$^{7}$,
M.~SEKIMOTO$^{10}$,
H.~SUGIMURA$^{7}$,
T.~TAKAHASHI$^{11}$,
H.~TAMURA$^{6}$,
K.~TANIDA$^{7}$ \and
W.~TRAUTMANN$^{1}$
}

\address{
$^{1}$ GSI Helmholtzzentrum f\"ur Schwerionenforschung GmbH, Plankstrasse 1., D-64291 Darmstadt, Germany\\
$^{2}$ Institute f\"ur Kernphysik, Johannes Gutenberg Universit\"at Mainz, Johann-Joachim-Becher-Weg 45, D-55099 Mainz, Germany\\
$^{3}$ RCNP, Osaka University, 10-1 Mihogaoka, Ibaraki, Osaka 567-0047, Japan\\
$^{4}$ Osaka Electro-Communication University, 18-8 Hatsumachi, Neyagawa, Osaka 572-8530, Japan\\
$^{5}$ KVI, Zernikelann 25, NL-9747 AA Groningen, The Netherlands\\
$^{6}$ Graduate School of Science, Tohoku University, 6-3 Aramaki Aza-Aoba, Aoba-ku, Sendai 980-8578, Japan\\
$^{7}$ Department of Physics, Kyoto University, Sirakawa Oiwake, Sakyou-ku, Kyoto 606-8502, Japan\\
$^{8}$ Department of Physics, Tokyo University, 7-3-1 Hongo, Bunkyo-ku, Tokyo 113-8654, Japan\\
$^{9}$ IPHC, Universit\'e Louis Pasteur, 23 rue du Loess, F-67037 Strasbourg cedex 2, France\\
$^{10}$ Department of Physics, Osaka University, 1-1 Machikaneyama, Toyonaka, Osaka 560-0043, Japan\\
$^{11}$ KEK, 1-1 Oho, Tsukuba, Ibaraki 305-0801, Japan\\
}



\maketitle

\begin{history}
\received{(received date)}
\revised{(revised date)}
\end{history}

\begin{abstract}
The HypHI collaboration aims to perform a precise hypernuclear spectroscopy with stable heavy ion beams
and rare isotope beams at GSI and FAIR in order to study hypernuclei at extreme isospin, especially neutron rich
hypernuclei to look insight hyperon-nucleon interactions in the neutron rich medium, and hypernuclear magnetic moments
to investigate baryon properties in the nuclei\cite{LOI,Sai06}. We are currently preparing for the first experiment with 
$^{6}$Li and $^{12}$C beams at 2~$A$~GeV to demonstrate the feasibility of a precise hypernuclear spectroscopy by 
identifying $^{3}_{\Lambda}$H, $^{4}_{\Lambda}$H and $^{5}_{\Lambda}$He\cite{Sai06,SaiProp}. The first physics experiment on these hypernuclei is planned for 2009.
In the present document, an overview of the HypHI project and the details of this first experiment will be discussed.
\end{abstract}

\section{Introduction to the HypHI project}
\subsection{Physics motivations}
The growing interest in hypernuclear physics is motivated by an increasing awareness of the importance to study hyperon-nucleon (YN) interactions and hyperon-hyperon (YY) interactions in order to fully understand baryon-baryon interactions under flavored SU(3)$_f$. The difficulty to study the YN and YY interactions by scattering experiments arises from the fact that it is impractical to achieve hyperon beams with proper energies and that no hyperon targets are available so far due to the short lifetime of hyperons ($\approx10^{-10}$~s). However, it has been demonstrated that, under certain conditions, hyperons can be bound into a nuclear system to form a hypernucleus which then can be used as a micro-laboratory to study YN and YY interactions.

\subsection{Production of hypernuclei}
\noindent Until recently, hypernuclei have mainly been produced and studied by meson or electron induced reactions with thick stable nuclear targets. In these experiments, a nucleon in the target nucleus is converted to a hyperon, thus the isospin of the produced hypernuclei is close to the one of the target nuclei. Moreover, due to the small momentum tranfert of the reaction, hypernuclear moments can not be accessed.\\
\noindent In order to overcome these difficulties, the HypHI collaboration proposes to produce exotic hypernuclei by heavy ion collisions with stable heavy ion beams and rare isotope beams (RI-beams). During the reaction, for a beam energy above the hyperon production threshold (1.6~$A$~GeV), an hyperon can be produced in the hot participant region and, due to the overlapping in the phase space of the hyperon and projectile spectator fragments, the hyperon may coalesce into the spectator fragment to form a hypernucleus.
Having the same velocity of the one of the projectile, the produced hypernuclei have a longer effective life time due to the Lorentz boost. Therefore, their decay takes place a few tens of centimeters behind the target. In this method, hypernuclei can be studied in flight.

\subsection{The phase 0 experiment at GSI}
\begin{figure}[th]
\centerline{\psfig{file=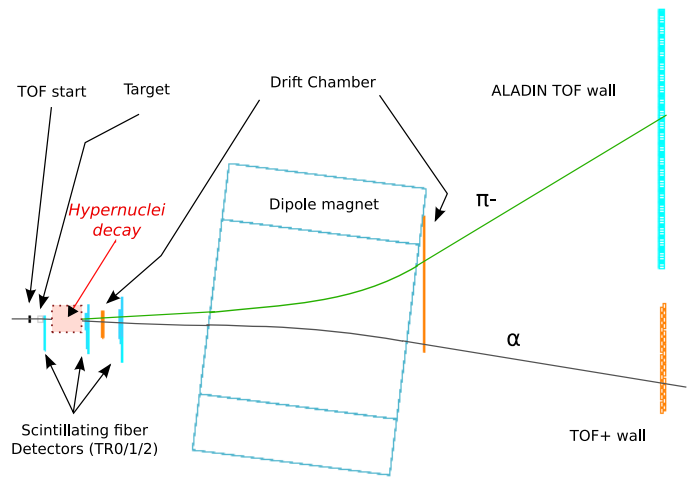,width=9cm}}
\vspace*{8pt}
\caption{A schematic drawing of the proposed experimental setup for the Phase 0 experiment at GSI.}
\label{fig:setup}
\end{figure}

\noindent The phase 0 experiment, planned for mid-2009 at GSI, is devoted to demonstrate the feasibility of a precise hypernuclear spectroscopy using heavy ion collisions. For this purpose, we will concentrate on studying light hypernuclei such as $^{3}_{\Lambda}$H, $^{4}_{\Lambda}$H and $^{5}_{\Lambda}$He produced by collisions of a $^{6}$Li and a $^{12}$C beam at 2~$A$~GeV with an intensity of $10^7$ particles per second on a thick carbon target (8~g/cm$^2$). These light hypernuclei will mainly be studied by looking into their respective mesonic weak decay (MWD) channels : $^{3}_{\Lambda}$H~$\rightarrow \pi^-+^{3}$He, $^{4}_{\Lambda}$H~$\rightarrow \pi^-+^{4}$He and $^{5}_{\Lambda}$He~$\rightarrow \pi^-+^{4}$He+p respectively.\\
The proposed setup for phase 0 experiment shown in Fig.~\ref{fig:setup} consists of the large acceptance dipole magnet ALADiN used to analyze the momenta of the different charged particles produced by the decay of hypernuclei, a time-of-flight (TOF) start detector in front of the target, three arrays of scintillating fiber tracking detectors (TR0, TR1 and TR2), the existing ALADiN TOF-wall for detecting $\pi^-$ particles and the currently under construction TOF+ wall used to detect the positively charged particles. In addition, two drift chambers will be placed before and after the magnet in order to increase the tracking efficiency and an other TOF-wall, along with the Large Area Neutron Detector (LAND), will be used respectively for the detection of $\pi^+$ and to investigate non-mesonic weak decay (NMWD) channels.

\noindent Due to the high beam intensity ($10^7$ particles per second), the trigger system for the data acquisition as well as the background reduction represent a major technical challenge in the success of the phase 0 experiment.
The phase 0 trigger decision system will be divided in three combined levels requiring three different conditions.
The first trigger level will be made from signals coming from the scintillating fiber tracking detectors. After signal processing in discriminator modules, LVDS logic signals are fed to new logic modules (VUPROM2) with FPGA and DSP chips, which are developed at GSI. These modules will then reject event by event the tracks corresponding to a primary vertex coming from reactions inside the target from those corresponding to a secondary vertex outside the target caused by hypernuclear decays. Monte Carlo simulations\cite{Geant} in the case of $^{4}_{\Lambda}$H reveal that the efficiency of this trigger is about $14~\%$ with a background reduction to $1.7~\%$. 

\begin{figure}[th]
\centerline{\psfig{file=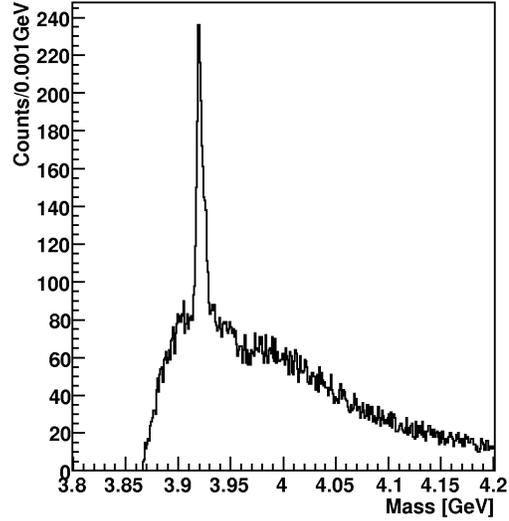,width=7.5cm}}
\vspace*{8pt}
\caption{Simulated invariant mass spectrum of $^{4}_{\Lambda}$H with background for 46900 hypernuclei events. A mass resolution of $\sigma=3.0$~MeV/$c^2$ is expected.}
\label{fig:mass}
\end{figure}

\noindent Another trigger requires the detection of a $\pi^-$ in the ALADiN TOF-wall whereas the third trigger requires a hit in the TOF+ wall corresponding to $Z=2$ particles which are the decay products of $^{3}_{\Lambda}$H, $^{4}_{\Lambda}$H and $^{5}_{\Lambda}$He. The calculated efficiencies for these two triggers in the case of $^{4}_{\Lambda}$H decay are found to be $28~\%$ for the second trigger and $94~\%$ for the third one with reduction of the background down to $15~\%$ and $14~\%$ respectively.
By combining these three different conditions, the efficiency for the the full trigger system for phase 0 experiment is expected to be of the order of $7~\%$ with a background reduction to $0.017~\%$.\\
 
\begin{table}[!h]
\tbl{Preliminary estimations of the rate for reconstructed hypernuclei events per week for Phase 0 experiment}
{\begin{tabular}{@{}ccc@{}} \toprule
Hypernucleus & Expected cross section ($\mu$b) & Reconstructed events per week \\
& & \colrule
$^{3}_{\Lambda}$H & 0.1 & 2.8 x $10^3$  \\ 
$^{4}_{\Lambda}$H & 0.1 & 2.6 x $10^3$ \\
$^{5}_{\Lambda}$He & 0.5 & 6.3 x $10^3$ \\
\end{tabular}}
\label{tab:trigger}
\end{table}

\noindent The invariant mass distribution obtained by Monte Carlo simulations for $^{4}_{\Lambda}$H is shown in Fig.~2. A mass resolution of $\sigma=3.0$~MeV/$c^2$ is expected for an estimated rate of reconstructed hypernuclei events of 2.6~x~$10^3$ events per week assuming a 0.1~${\mu}b$ cross section (Tab.~1).

\section{Summary and outlook}
In order to achieve comprehensive understanding of the baryon-baryon interaction, the study of hypernuclei is essential. The HypHI project at GSI and FAIR aims to perform a precise hypernuclear spectroscopy with heavy ion beams in order to study hypernuclei at extreme isospin and to measure hypernuclear magnetic moments for the first time. A pilot experiment, phase 0, scheduled for mid-2009 is devoted to demonstrate the feasibility of such a spectroscopy by looking into the mesonic weak decay (MWD) channels of light hypernuclei such as $^{3}_{\Lambda}$H, $^{4}_{\Lambda}$H and $^{5}_{\Lambda}$He.\\
After the success of this pilot experiment, we would like to extend our investigations to heavier hypernuclei especially towards the hypernuclear proton drip-line by using rare isotope beams delivered by the fragment separator FRS at GSI. Furthermore, with the future facility at GSI (FAIR), we will be able to perform experiments on neutron rich hypernuclei using neutron rich RI-beams delivered by the super-FRS. Finally, with beam energies of the order of 20~$A$~GeV available at FAIR facility, one of the final goal of the HypHI project, namely measurements of hypernuclear moments, will be reached. For this purpose, we want to develop a hypernuclear separator made of superconducting bending magnets to be able to separate the produced hypernuclei according to their respective magnetic rigidity. 

\section*{Acknowledgements}
I would like to thank the organizers of the Franco-Japanese Symposium 2008 for giving me the opportunity to present our work in such a pleasant atmosphere.\\

\noindent \it{The HypHI project is supported by the Helmholtz association as Helmholtz-University Young Investigators Group VH-NG-239 and by the German Research Foundation (DFG) with a contract number SA 1696/1-1.}

\end{document}